\documentclass[preprint,nofootinbib]{revtex4-1}

\usepackage{graphicx}
\usepackage{amssymb}
\usepackage{amsmath}
\usepackage{natbib}

\DeclareMathAlphabet{\mathpzc}{OT1}{pzc}{m}{it}

\begin{document}


%
%

\title{Topological supersymmetry breaking:
The definition and stochastic generalization of chaos and the limit of applicability of statistics}

\author{Igor V. Ovchinnikov} 
\email{iovchinnikov@ucla.edu}
\author{Robert N. Schwartz}
\author{Kang L. Wang}
\affiliation{Electrical Engineering Department, University of California at Los Angeles, \\
Los Angeles, CA 90095 USA}

\begin{abstract}
The concept of deterministic dynamical chaos has a long history and is well established by now. Nevertheless, its field theoretic essence and its stochastic generalization have been revealed only very recently. Within the newly found supersymmetric theory of stochastics (STS), all stochastic differential equations (SDEs) possess topological or de Rahm supersymmetry and stochastic chaos is the phenomenon of its spontaneous breakdown. Even though the STS is free of approximations and thus is technically solid, it is still missing a firm interpretational basis in order to be physically sound. Here, we make a few important steps toward the construction of the interpretational foundation for the STS. In particular, we discuss that one way to understand why the ground states of chaotic SDEs are conditional (not total) probability distributions, is that some of the variables have infinite memory of initial conditions and thus are not "thermalized", \emph{i.e.}, can not be described by the initial-conditions-independent probability distributions. As a result, the definitive assumption of physical statistics that the ground state is a steady-state total probability distribution is not valid for chaotic SDEs.
\end{abstract}

\maketitle

\section{Introduction}

Time evolution of a dynamical system (DS) is specified directly by its equations of motion that in the general case do not follow from the least action principle. This is the most general class of continuous-time dynamical models as compared, \emph{e.g.}, to classical or Hamilton dynamics  known in the DS theory as conservative models. As a result of this generality, DS theory has the widest applicability ranging from the social and biological sciences to astrophysics.

A natural DS always experiences a stochastic influence of external noise coming from its environment. In other words, natural DSs are described by stochastic (partial) differential equations (SDEs).\footnote{We will not consider here quantum dynamics - yet another large category of dynamics in the natural world. Nevertheless, the supersymmetric approach to SDEs can also be useful for quantum systems such as spin quantum systems.\cite{Grindel}} Most of the efforts in the DS theory, however, were directed toward idealistic deterministic dynamics. In particular, even though some important insights on the interplay between noise and chaotic behavior have already been established, \cite{Kapitaniak} it is still an open question what is a mathematically rigorous stochastic generalization of the concept of dynamical chaos - a very fundamental dynamical phenomenon \cite{review} with over a century long history.\cite{Rue14} 

Many important properties of deterministic chaos are well now established. For example, it is a common knowledge that the onset of chaos has features of a phase transition such as universality. \cite{Univ_Chao} It is also known that chaos has a lot to do with topology: at the transition into deterministic chaos fractal attractors, which are not well-defined topological manifolds, show up. The connection to topology actually goes deeper: fractal attractors in 3D consist of unstable periodic orbits with nontrivial linking numbers laying at the heart of "chaos topology". \cite{ChaosTopology}

A related topic is the mysterious dynamical long-range order that reveals itself through such well-established phenomena and concepts as the 1/f noise, the butterfly effect (or the sensitivity to initial conditions), and the flicker noise and/or the power-law statistics of instantonic processes in the seemingly unrelated natural DSs on all observable scales. The later include the neurodynamical avalanches in brain, \cite{brain}
earthquakes, \cite{earthquakes}
solar flares, \cite{SolarFlares}
biological \cite{biologicalevolution}
and celestial \cite{celestialevolution}
evolutions,
financial markets, \cite{FinancialMarkets}
glasses, \cite{glasses}
various nanometer scale devices \cite{Nano}
and many, many others. The understanding of the mathematical origin of this ubiquitous dynamical long-range order is an important outstanding problem of modern science.

A consistent theory promising to connect all these dots (\emph{i.e.}, to establish the topological nature of chaos, to explain the origin of the dynamical long-range order, and to work just as well for stochastic DSs) has been very recently proposed by one of us. \cite{Mine,Mine0,Mine1,Mine2} Within this supersymmetric theory of SDEs or simpler of stochastics (STS), all SDEs possess topological or de Rahm supersymmetry, stochastic chaotic behavior should be associated with its spontaneous breakdown, and the dynamical long-range order seems to be the consequence of the Goldstone theorem.

The STS is free of approximations and thus is technically solid. On the other hand, it is still missing an interpretational foundation that would provide the key ingredients of the theory with a clear physical meaning. In this paper, we make a few steps toward the construction of the interpretational basis of the STS. We show that the existence of the topological supersymmetry in all SDEs is the algebraic representation of the fact that smooth dynamics respects the concept of boundary. We also discuss the physical meaning of "chaotic" or non-supersymmetric ground states that are conditional probability distributions. \footnote{The DS theory predecessor of such ground states is the Sinai-Ruelle-Bowen conditional probability function. \cite{ErgodicTheoryOfChaos}} We argue that such ground states are representatives of the situation when a DS does not reach its thermodynamic equilibrium because it fails to "thermalize" some of its variables, in which it has infinite memory of the initial conditions and/or perturbations. In these variables, the ground state is not a "distribution" and in order to make probabilistic sense out of such a wavefunction someone or something must know the values of these variables.

\section{Random "Step-Like" Dynamics}

The story of the STS has two sides: the pathintegral or field-theoretic side and the DS theory side. \cite{Mine} Here, we choose the second approach because it is advantageous when it comes to the physical interpretation of various objects. 

Let us begin with the discussion of a deterministic step-like DS, \emph{i.e.}, a $D$-dimensional topological (and orientable) manifold called the phase space, $X$, and the map of $X$ onto itself
\begin{eqnarray}
&M:X\to X. \label{DSStepLikeTime}
\end{eqnarray}
The map has a very clear meaning: it takes a DS initially at $x_{in}$ and places it at $x_{out} = M(x_{in}), x_{in}, x_{out} \in X$.

In this paper, we are interested in maps defined by the "physical" SDEs, \emph{i.e.}, by the SDEs with "smooth enough" flow vectors fields and $e$'s (see the next section). A map defined by a physical SDE (with a fixed noise configuration) is a diffeomorphism. In particular, it is differentiable and invertible. This is the class of maps that we limit our consideration to.

Unlike in the conventional approaches to stochastics, in the STS, the Hilbert space is the entire exterior algebra of the phase space and not only the space of the total probability distributions that in the coordinate-free setting are the top differential forms. This is actually a mathematical necessity that follows, in particular, from the fact that if one considers only the space of top differential forms, then the Witten index of the SDE, that represents (up to a topological factor) the partition function of the stochastic, would not exist (see discussion at the end of Sec.3 below). On the other hand, the partition function of the noise is a very fundamental and in a sense definitive object for the formulation of the model, so that the theories that do not contain the representative of this object can not be viewed as complete. 

The Hilbert space now is the space of the differential forms:
\begin{eqnarray}
&\psi^{(k)} = (k!)^{-1} \psi^{(k)}_{i_1...i_k} dx^{i_1} \wedge ... \wedge dx^{i_k} \in \Omega^{k}(X),\label{DForms}
\end{eqnarray}
where $\psi^{(k)}_{i_1...i_k}$ is an anti-symmetric contravariant tensor of rank $k$, $\wedge $ is the so-called wedge product, the essence of which is the antisymmetrization, e.g., $dx^1\wedge dx^2 = - dx^2\wedge dx^1 = dx^1 \otimes dx^2 - dx^2 \otimes dx^1$, and $\Omega^{k}(X)$ is the linear space of all differential forms of degree $k$, \emph{i.e.}, $k$-forms.

One possible interpretation of such wavefunctions was given in Ref.\cite{Mine}a. There, the differential forms are proposed to be viewed as the generalized probability distributions (GPD) in the coordinate-free setting.\footnote{This is the picture that we adopt here, even though we believe that the physical meaning of wavefunctions and the information contained in them may be richer.} This can be clarified as follows. The differential forms are naturally coupled by integration to the (lower-)dimensional submanifolds in $X$ that are also known as chains:
\begin{eqnarray}
&p(c_k) = \int_{c_k}\psi^{(k)} \in \mathbb{R}^1,\label{expectationvalue}
\end{eqnarray}
where $c_k\in C_k(X)$ is a k-(dimensional-)chain and $C_k(X)$ is the group of all k-chains. In this manner, a top differential form, $\psi^{(D)}$, has the meaning of the total probability distribution and $p(c_D) = \int_{c_D}\psi^{(D)}$ is the probability of finding the DS within the volume $c_D\in C_D(X)$. For $k<D$, the wavefunction can be interpreted as a "lower-dimensional" probability distribution, \emph{i.e.}, they are distributions only in those dimensions in which they have differentials. The variables in the dimensions that have no differentials must be viewed as the conditional parameters for the distribution. In other words, such a wavefunction can be viewed as a conditional probability distribution in a generalized (or local) sense.\footnote{If the generalized probability distribution at a given point is negative, this can be remedied by changing the orientation of (some of) the local coordinates at this point. Globally, however, this is not possible in the general case.} This interpretation of the wavefunction of degree 2 in a three-dimensional phase space is illustrated in Fig. 1b.

The conventional (or global) conditional probability distributions can be defined in the coordinate-free setting for $X=\mathbb{R}^D$ as, $P_{tot}=P_{cond}\wedge P_{marg}$, where $P_{tot}(x)dx^1\wedge ...\wedge dx^D$, $P_{cond}(x^1...x^k|x^{k+1}...x^D)dx^1\wedge ...\wedge dx^k $, $P_{marg}(x^{k+1}...x^D)dx^{k+1}\wedge...\wedge dx^D$, and $P_{marg}=\int P_{tot}(x)dx^1\wedge ...\wedge dx^k$. These objects carry no additional information about the system as compared to the information contained in $P_{tot}$ itself. On the contrary, the wavefunctions of STS may contain more information than the total probability distribution associated with it. \footnote{Getting a little bit ahead, the bra-ket combination of each eigenstate is a top differential form that can be interpreted as the total probability distribution associated with this eigenstate.} The situation is akin to that in quantum mechanics where the wavefunction does not only represent the total probability distribution (or rather the square root of that) but also contains the quantum mechanical phase.

\begin{figure}[h]
\centering
\includegraphics[height=8.0cm, width=9.5cm]{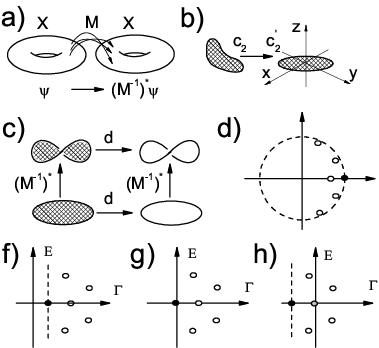}
\caption{{\bf (a)} A DS can be defined through a map, $M$, of the phase space, $X$, to itself. The differential forms on $X$, $\psi$'s, evolve by the pullback, $(M^{-1})^*$, of the inverse diffeomorphism, $M^{-1}$, of the evolution, $M$. ({\bf b}) The differential forms of degree less then the dimensionality of $X$ can be interpreted as generalized or conditional probability distributions. In the example given, a 2-chain, $c_2$, in a three-dimensional phase space can be "straightened out", $c_2\to c'_2 = f(c_2)$, by the appropriate coordinate transformation (arrow), that induces the corresponding transformation on the wavefunction (a 2-form), $\psi\to\psi'=(f^{-1})^*\psi\in\Omega^2(X)$. The quantity in Eq.(\ref{expectationvalue}) is $\int_{c_2}\psi = \int_{c'_2}\psi' = \int_{(x,y)\in c'_2}\psi'_{xy} dx\wedge dy$ can be viewed as a probability to find $x$ and $y$ within $c'_2$ given that $z=0$, and $\psi'_{xy}$ must be interpreted as a (local) conditional probability distribution, $P(xy|z)$. ({\bf c}) The exterior derivative, $\hat d$, commutative with pullbacks, $(M^{-1})^*$, acts on the Poincar\'e duals of chains as a boundary operator. The commutativity diagram and/or the presence of the topological supersymmetry, $\hat d$, can be interpreted as though the smooth dynamics respects the concept of boundary. ({\bf d}) The spectrum of the generalized transfer operator (finite-time stochastic evolution operator) from the spectral theorems of the DS theory: the largest magnitude eigenvalue is real. ({\bf f})-({\bf h}) The three corresponding possible spectra of the Fokker-Planck Operator. Situation ({\bf f}) is ruled out at least for DSs with non-zero Euler characteristic of the phase space. The other two situations correspond to unbroken (g) and broken (h) topological supersymmetry. Black dots represent ground states. The dot with grey filling at the origin in (h) represents the $\hat d$-symmetric state(s) ("ergodic zero(s)") that are no longer ground state(s). \label{Figure1}}
\end{figure}

The law for the "temporal" evolution of the wavefunctions induced by map (\ref{DSStepLikeTime}) can be found by taking the coordinate-free object from Eq.(\ref{DForms}), writing it down in terms of $x_{in}$, and then rewriting it in terms of $x_{out}$ using $x_{in} = M^{-1}(x_{out})$ as a formal coordinate transformation. This leads to what is known as the pullback induced by $M^{-1}$:
\begin{subequations}
\begin{eqnarray}
&(M^{-1})^*: \Omega^k(X) \to \Omega^k(X),
\end{eqnarray}
the coordinate version of which is
\begin{eqnarray}
&((M^{-1})^*\psi^{(k)})_{i_1...i_k}(x_{out}) = \nonumber \\
&\psi^{(k)}_{\tilde i_1...\tilde i_k}(M^{-1}(x_{out}))  (TM^{-1})^{\tilde i_1}_{i_1}(x_{out})...(TM^{-1})^{\tilde i_k}_{i_k}(x_{out}),\label{pullback}
\end{eqnarray}
\end{subequations}
with $(TM^{-1})^{\tilde i}_{i}(x_{out}) = \partial (M^{-1}(x_{out}))^{\tilde i}/\partial x^{i}_{out}$ being the tangent map $TM^{-1}: T_{x_{out}}X \to T_{x_{in}}X$ from the tangent space, $TX$, at $x_{out}$ to that at $x_{in}$.

So far, we only changed the coordinate representation of a coordinate-free object - the differential form of interest. The actual change of the differential form comes with what could be called the "time shift". In order to be able to compare what we had before with what we get after the time evolution we make the formal substitution $x_{out}\to x_{in}$ in Eq.(\ref{pullback}). In other words, we identify the phase space after the evolution with the phase space before it. As a result, the time evolution of a differential form is the pullback induced by the diffeomorphism, $M^{-1}$.

The next step is the generalization of the discussion to stochastic dynamics. Consider maps, $M(\xi)$, that depend on some stochastic variables (noise), $\xi$, with a normalized probability distribution $P(\xi)$, $\int P(\xi)d\xi = 1$. We introduce now the generalized transfer operator (GTO):\cite{Ruelle}
\begin{eqnarray}
&\hat{\mathpzc{M}}:\Omega^k(X) \to \Omega^k(X),
\end{eqnarray}
which is the stochastically averaged pullback
\begin{eqnarray}
&\hat{\mathpzc{M}} = \int (M^{-1}(\xi))^* P(\xi)d\xi\equiv\langle M^{-1}(\xi))^*\rangle_{noise}.\label{GTO}
\end{eqnarray}
The averaging here is legitimate because pullbacks are linear operators on a linear Hilbert space, which is the exterior algebra of $X$, $\Omega(X)= \bigoplus_{k=0}^D\Omega^k(X)$, \emph{i.e.}, the linear space of all differential forms of all degrees. At this moment, it may become suspicious that highly nonlinear dynamics can be described by linear operators. The same is true for any quantum theory, in which the evolution is a linear (unitary) operation on a linear infinite-dimensional Hilbert space, whereas the underlying classical equations of motion can be in finite-dimensions and highly nonlinear. 

The price we pay for this "linearization" is that the Hilbert space is infinitely more dimensional than the DS itself. One may now wonder whether this price might be too high and the introduction of the Hilbert space is an unnecessary complication. The answer is no because the probabilistic description is a necessity for the stochastic dynamics and/or the ergodic approach to deterministic dynamics.

It is already at this point that we can establish the existence of topological supersymmetry. Indeed, let us now introduce the exterior derivative, also known as de Rahm operator, $\hat d = dx^i\wedge \partial/\partial x^i, \hat d: \Omega^k(X) \to \Omega^{k+1}(X)$. One of its properties is the commutativity with pullbacks. In particular, $(M^{-1}(\xi))^*\hat d = \hat d (M^{-1}(\xi))^*$ for each configuration of noise, $\xi$, in Eq.(\ref{GTO}) so that the GTO also commutes with $\hat d$:
\begin{eqnarray}
[\hat d, \hat{\mathpzc{M}}] = 0.\label{dUCommute}
\end{eqnarray}
By analogy with quantum mechanics,\footnote{The continuous-time version of the theory we have so far is related to topological quantum mechanics that we discuss below.} the GTO must be identified as a finite-time evolution operator and $\hat d$ as a conserved quantity. From a group theoretic point of view, $\hat d$ is a generator of the Abelian one-parameter group, $\mathcal{G} = \{\hat g_s = e^{s\hat d}| s\in {R}^1\}$, of a continuous symmetry of the DS:
\begin{eqnarray}
(\hat g_{s})^{-1}\hat{\mathpzc{M}}\hat g_{s} = \hat{\mathpzc{M}}.
\end{eqnarray}
The above equality can be easily proved with the use of Eq.(\ref{dUCommute}), the nilpotency property, $\hat d^2 = 0$, and $e^{s\hat d} = \hat 1 + s\hat d$, which follows from the nilpotency property.

Just as for any other symmetry, the eigenstates of the GTO of the same eigenvalue make up irreducible representations of $\hat d$-symmetry. There are only two types of irreducible representations of $\hat d$-symmetry - the one-dimensional $\hat d$-symmetric singlets and two-dimensional non-$\hat d$-symmetric bosonic-fermionic (BF) pairs (multiplets).\footnote{In the high-energy physics language, the states with even and odd number of fermions (ghosts in our case, see below) are referred to as bosonic and fermionic states respectively.} In the eigenstates' basis, the exterior derivative has the following form:
\begin{eqnarray*}
\hat d = {diag}\left(0,...,0,
\left(\begin{array}{cc}
0 & 1 \\
0 & 0 \end{array}\right),
\left(\begin{array}{cc}
0 & 1 \\
0 & 0 \end{array}\right),\dots\right),\label{explicitd}
\end{eqnarray*}
with zeros corresponding to the $\hat d$-symmetric singlets and the Jordan blocks corresponding to the non-$\hat d$-symmetric pairs.

Following Ref.\cite{WittenForms}, we can now turn to the field theoretic representation of the theory. This is done by the formal introduction of the anticommuting fields that in the pathingeral formulation of the theory can be called the Fadeev-Popov ghosts. The ghosts are formal notation for the differentials, $\chi^i \equiv dx^i \wedge$. In this notation, the wavefunctions in Eq.(\ref{DForms}) become
\begin{eqnarray}
&\psi^{(k)}(x,\chi) = (k!)^{-1} \psi^{(k)}_{i_1...i_k} (x) \chi^{i_1} ... \chi^{i_k},\label{wavefunctions}
\end{eqnarray}
with $\hat d=\chi^i\partial/\partial x^i$. It is clear now that $\hat d$ is a supersymmetry, \emph{i.e.}, a symmetry that mixes the commuting ($x^i$'s) and the anticommuting ($\chi^i$'s) fields.

The presence of this supersymmetry can be understood as follows. The exterior derivative is the algebraic representative of the concept of boundary. For example, it lies at the heart of Stoke's theorem relating the integration over the boundary of a chain to that over its interior: $\int_{\partial c_k} \psi^{(k-1)} = \int_{c_k} \hat d \psi^{(k-1)}$, where $\partial c_k$ is the boundary of $c_k\in C_k(X)$. 

Let us recall now the concept of the Poincar/'e dual. One of the versions of Poincar\'e duality states that for any chain, $c_k\in C_{k}(X)$, there is a differential form, $\overline{c_k}\in \Omega^{D-k}(X)$, called Poincar\'e dual, such that $\int_{c_k}\psi^k = \int_{X}\overline{c_k}\wedge \psi^k$ for any $\psi^{k}\in\Omega^k(X)$. The differential form $\overline{c_k}$ is a delta-functional distribution on the chain, $c_k$, with the differentials/ghosts in the transverse directions. 

Stoke's theorem can be used to show that $\hat d$ acts (up to a sign) on the so-called Poincar\'e duals of chains as the boundary operator would have acted on the corresponding chains, $\hat{d}\overline{c}=\overline{\partial c}$ (see Fig.\ref{Figure1}c). For example, for the segment, $c = \{ x,y | x_1>x>x_2, y=y_0\}$ on the $x-y$ plane, the Poincar\'e dual is, $f(x)\delta(y-y_0)dy$, where $f(x)=1$ for $x_1>x>x_2$ and zero otherwise. It is easy to see that, $\hat d \overline{c} = \delta(x-x_1)dx\wedge \delta(y-y_0)dy - \delta(x-x_2)dx\wedge \delta(y-y_0)dy =  \overline{c_1} - \overline{c_2}$ , where $c_{1,2}=\{ x,y | x=x_{1,2}, y=y_0\}$ are the two boundaries of chain, $c$. In other words, $\hat d$ can be viewed as an algebraic representative of the concept of boundary.

The presence of this supersymmetry in all natural DSs is an algebraic version of the statement that smooth dynamics respects the concept of boundary. In other words, the action of diffeomorphisms on chains is commutative with the procedure of taking the boundary of these chains. The cohomoliogical version of this statement is that the pullback induced by any diffeomorphism is commutative with the exterior derivative as is visually exemplified in Fig.\ref{Figure1}c. The concept of boundary is very fundamental, which explains why all natural DSs must possess the topological supersymmetry.

Allow us now to digress for a moment on the history of this supersymmetry. It was first introduced in the context of stochastic DSs by Parisi and Sourlas \cite{ParSour} who derived the most general stochastic quantization procedure for Langevin SDEs as compared to the already existing formalism at that time known as the Martin-Siggia-Rose procedure. \cite{MartinSR} It was found that theories that emerge from the stochastic quantization of Langevin SDEs are (N=2) supersymmetric. Later, this supersymmetry was related to algebraic topology \cite{WittenForms} and yet later identified as the gauge-fixing Betti-Route-Stora-Tuytin (BRST) symmetry \cite{IdentifBRST} and topological supersymmetry ($\mathcal Q$-symmetry) \cite{Labastida} - a unique attribute of cohomological or Witten-type topological field theories (TFT's). \cite{TFTBook} The simplest nontrivial member of this family of models is the topological quantum mechanics. \cite{Labastida}

Langevin SDEs, however, are not a sufficiently large class of DSs to identify chaos with the $\mathcal Q$-symmetry breaking.\footnote{Furthermore, $\mathcal Q$-symmetry is never broken for Langevin SDEs as the Fokker-Planck eigenvalues (see below) are all real and non-negative.\cite{Mine}b} Thus, the generalization of the Parisi-Sourlas method to all SDEs was needed. At this, on the formal application of the Parisi-Sourlas method to other classes of SDEs, pseudo-Hermitian models show up (see, \emph{e.g.}, Refs. \cite{Kurchan} and \cite{ClassicalMechanics,ClassicalMechanics2} for Kramers' equation and classical/conservative dynamics, respectively). Therefore, even though it was known previously \cite{Gaw} that the Parisi-Sourlas quantization procedure formally applied to a general form SDE leads to a pathintegral representation of a model with a $\mathcal Q$-exact action, \cite{Gaw} the generalization became possible only after the rigorous formulation of the theory of the pseudo-Hermitian operators. \cite{Mostafa} This generalization \cite{Mine,Mine1,Mine2} showed that the approximation-free stochastic quantization of any SDE leads to a pseudo-Hermitian model with $\mathcal Q$-symmetry. 

Note, however, the STS cannot be recognized as a full-fledged cohomological theory because in conventional cohomological theories one limits ones interest only to supersymetric states, whereas of primary interest in the STS are the ground states that for chaotic models are not symersymmetric. 

\section{Stochastic Dynamics}

For continuous-time stochastic DSs, dynamics can be defined by the following SDE:
\begin{eqnarray}
\dot{x}^i(t) = F^i(x(t)) + (2\Theta)^{1/2}e^i_a(x(t))\xi^a(t)\equiv\mathcal{F}(t),\label{SDE}
\end{eqnarray}
where $F(x)\in T_{x}X$ is the so called flow vector field from the tangent space of $X$, $\xi$'s are parameters of the noise, $e^i_a(x)$ are vielbeins so that the metric on $X$ is $g^{ij}=e^i_a(x) e^j_a(x)$, and $\Theta$ is a parameter representing the intensity or temperature of the noise. The limit, $\Theta\to0$, is the deterministic limit where the SDE (\ref{SDE}) becomes an ordinary differential equation. 

For Gaussian white noise, $\langle\xi^a(t)\xi^b(t') \rangle_{noise}=\delta^{ab} \delta(t-t')$, the GTO corresponding to the temporal evolution of duration $t$ has the form of the finite-time evolution operator,
\begin{eqnarray}
\hat{\mathpzc{M}}_t = e^{-t\hat H}, \label{GTOTopQuant}
\end{eqnarray}
which corresponds to the infinitesimal temporal evolution given by the generalized Fokker-Planck (FP) equation for differential forms,
\begin{eqnarray}
\partial_t \psi(t) = - \hat H\psi(t), \label{GenFPEq}
\end{eqnarray}
The stochastic evolution operator or the generalized FP operator is defined as,
\begin{eqnarray}
\hat H = \hat{\mathcal{L}}_{F} - \Theta \hat{\mathcal{L}}_{e_a}\hat{\mathcal{L}}_{e_a} = [\hat d, \hat {\bar d}].\label{FPHam}
\end{eqnarray}
Here,\footnote{The differentiation over the fermionic variables can be defined as: $(\partial/\partial \chi^i) \chi^{j_1}...\chi^{j_k} = \sum_{l=1}^k(-1)^{l-1}\chi^{j_1}...\chi^{j_{l-1}}\delta^{j_l}_i\chi^{j_{l-1}}...\chi^{j_k}$.} $\hat{\mathcal{L}}_{F} = F^i \partial/\partial x^i + F^i_{'j} \chi^j \partial/\partial\chi^i = [\hat d, \hat \imath_F]$ is the Lie derivative along the flow with $\hat \imath_F \equiv \partial/\partial\chi^i F^i$ being the interior multiplication by $F$,\footnote{Its action on, \emph{e.g.}, a 2-form is, $\hat \imath_F (1/2!)\psi^{(2)}_{ij}\chi^i\chi^j = (1/2)(\psi^{(2)}_{ij}F^i\chi^j - \psi^{(2)}_{ij}\chi^i F^j) =  \psi^{(2)}_{ij}F^i\chi^j$.}  $\hat{\mathcal{L}}_{e_a}\hat{\mathcal{L}}_{e_a}$ can be called the diffusion Laplacian, and $\hat {\bar d} = \hat \imath_F - \Theta \partial/\partial\chi^i e^i_a \hat{\mathcal{L}}_{e_a}$ is the operator identified sometimes as the current operator. 

In the above formulas, the square brackets denote a bi-graded commutator, which is defined as $[\hat a, \hat b] = \hat a \hat b - (-1)^{I(\hat a)I(\hat b)}\hat b \hat a$, where $I$'s are the ghost degrees of the operators, \emph{i.e.}, the difference in the numbers of $\chi$'s and $\partial/\partial\chi$'s. For example, $I(\hat d) = - I(\hat {\bar d}) = 1$, so that $[\hat d, \hat {\bar d}]$ in Eq.(\ref{FPHam}) is actually an anticommutator.

The easiest way to derive Eqs.(\ref{FPHam}) and (\ref{GTOTopQuant}) is as follows (see Ref.\cite{Mine2} for details). First, for a fixed configuration of noise, the pullback, $(M^{-1}_{t0})^*= M^*_{0t}$, induced by the diffeomorphism, $M^{-1}_{t0}=M_{0t}$, of the inverse SDE-defined evolution during the time interval, $t>\tau>0$, can be given as:
\begin{eqnarray}
M^*_{0t} = {\mathcal T} e^{-\int_{0}^t d\tau \hat{\mathcal L}_{{\mathcal F}(\tau)}} = 1 - \int_{0}^t d\tau \hat{\mathcal L}_{{\mathcal F}(\tau)} + \int_{0}^t d\tau_1 \int_{0}^{\tau_1} d\tau_2 \hat{\mathcal L}_{{\mathcal F}(\tau_1)} \hat{\mathcal L}_{{\mathcal F}(\tau_2)} ...\label{Taylor}
\end{eqnarray}
Here, $\hat{\mathcal L}_{{\mathcal F}(\tau)}$ is the Lie derivative along the r.h.s. of Eq.(\ref{SDE}) and $\mathcal T$ is the operator of chronological ordering needed because at different time moments $\hat{\mathcal L}_{{\mathcal F}(\tau)}$'s do not commute. The above formula follows from the mathematical meaning of Lie derivative, which is the infinitesimal pullback of the flow along a vector field.

The stochastic evolution operator can be defined as:
\begin{eqnarray}
\hat H = \lim_{\Delta t\to 0} \Delta t^{-1}\left(1 - \left\langle M^*_{0\Delta t}\right\rangle_{noise} \right),
\end{eqnarray}
which follows from Eqs.(\ref{GTOTopQuant}) and (\ref{GTO}). Using now Eq.(\ref{Taylor}), the linearity of the Lie derivative in its argument, $\hat{\mathcal L}_{{\mathcal F}(\tau)} = \hat{\mathcal L}_{F} - (2\Theta)^{1/2}\xi^a(\tau)\hat{\mathcal L}_{e_a}$, and the standard correlators of the Gaussian white noise, \emph{i.e.}, $\left\langle \xi^a(\tau)\right\rangle_{noise} =0$ and the second order $\delta$-functional correlator defined earlier, one readily arrives at Eq.(\ref{FPHam}).\footnote{Note that the diffusion term will acquire the factor $2 \lim_{\Delta t\to 0} \Delta t^{-1} \int_0^{\Delta t}d\tau_1 \int_0^{\tau_1}d\tau_2\delta(\tau_1-\tau_2)$. This factor equals unity because the upper limit of the integration over $\tau_2$ is right at the peak of the $\delta$-function, which is a symmetric function of its argument so that, $\int_{0}^{\tau_1}d\tau_2\delta(\tau_1-\tau_2)=1/2$.}  

Eqs. (\ref{GTOTopQuant}) and (\ref{FPHam}) have a clear physical meaning. In the deterministic limit, $T\to0$, the evolution operator operator consists only of the Lie derivative, which by its definition is an infinitesimal pullback so that its exponentiation in Eq.(\ref{GTOTopQuant}) leads directly to a finite-time pullback. The stochastic noise, in turn, introduces diffusion. As it should, diffusion is represented by a member of the family of Laplace operators.

For top differential forms, the Lie derivative has the following form, ${\mathcal L}_{G}\psi^{(D)}=\partial/\partial x^i G^i \psi^{(D)}$, where $G\in TX$ is an arbitrary vector field on $X$. Thus, Eq.(\ref{GenFPEq}) is:
\begin{eqnarray}
\partial_t \psi^{(D)} = - \left( \frac\partial{\partial x^i} F^i - \Theta \frac\partial{\partial x^i} e^i_a\frac\partial{\partial x^i} e^j_a \right)\psi^{(D)},\label{convFPEq}
\end{eqnarray}
which is nothing else but the conventional FP Eq. for the total probability distribution. This Eq. corresponds to what is known in the classical theory of SDEs as the Stratonovich interpretation of stochastics. Note, however, that the SDE (\ref{SDE}) and the choice of the Gaussian white noise uniquely define the stochastic evolution operator and there is no need for any additional "interpretations".

Yet another famous physical Eq. that Eq.(\ref{GenFPEq}) has as a special case is the Liouville equation for Hamilton DSs. Indeed, in the deterministic limit ($\Theta=0$) and for the Hamilton flow vector field on $\mathbb{R}^{2D}$, $F^{x^i} = \partial H/\partial p_i$ and $F^{p_i} = - \partial H/\partial x^i$, with $i=1,...,D$, Eq.(\ref{convFPEq}) reduces further to:
\begin{eqnarray}
\partial_t \psi^{(2D)} = - \left( \frac{\partial H}{\partial p_i} \frac\partial{\partial x^i} - \frac{\partial H}{\partial x^i} \frac\partial{\partial p_i}  \right)\psi^{(2D)},
\end{eqnarray}
where we also used that $\partial^2 H/\partial x^i\partial p_i = \partial^2 H/\partial p_i\partial x^i$, which allows to drag the flow vector field through the differentials to the left. 

The stochastic evolution operator (\ref{FPHam}) is $\hat d$-exact, \emph{i.e.}, it is of the form of a bi-graded commutator.\footnote{The diffusion Laplacian and the Lie derivative are $\hat d$-exact separately, on their own.} This is a unique feature of cohomological field theories,\cite{TFTBook} for which the energy-momentum tensor (that reduces in the case of (topological) quantum mechanics to the Hamiltonian only) is $\mathcal Q$-exact.\footnote{As we already mentioned, the pathintegral version of $\hat d$ is called $\mathcal Q$, so that in high-energy physics terms $\hat d$-exact is $\mathcal Q$-exact.} This form of the evolution operator automatically suggests that $\hat d$ is a symmetry of the model, $[\hat d, \hat H] = 0$, because $[\hat d,[\hat d, \hat X]]=0, \forall \hat X$.

As a pseudo-Hermitian operator, the stochastic evolution  operator together with its eigensystem has the following properties. \cite{Mostafa} The eigenstates constitute a complete bi-orthogonal basis, $\hat H |\alpha\rangle = \mathcal{E}_\alpha |\alpha\rangle, \langle \alpha | \hat H = \langle \alpha | \mathcal{E}_\alpha$ such that $\langle \alpha| \beta \rangle \equiv \int_X\bar \alpha(x)\wedge \beta(x) = \delta_{\alpha \beta}$, and the resolution of unity on $\Omega(X)$ is $\hat {\mathbf{1}}_{\Omega} = \sum\nolimits_{\alpha}|\alpha\rangle\langle\alpha|$. Note that unlike for Hermitian evolution operators as in quantum theory, the bras and kets of the psuedo-Hermitian stochastic evolution operator non-trivially relate to each other. In particular, $\bar\alpha(x) \ne \star \alpha(x)^*$, where $\star$ denotes the Hodge conjugation.

Consider now an eigenstate, $\hat H|\alpha\rangle = \mathcal{E}_\alpha |\alpha\rangle$, with a non-zero eigenvalue, $\mathcal{E}_\alpha\ne 0$. Now, if $\hat d |\alpha\rangle\ne 0$, then due to $[\hat H,\hat d]=0$ we have a BF pair of eigenstates (see the discussion of Eq.(\ref{explicitd})) of the same eigenvalue: $|\alpha\rangle$ and $\hat d |\alpha\rangle$. The other possibility is $\hat d |\alpha\rangle = 0$. In this case, $|\alpha\rangle = \hat d |\alpha'\rangle$ with $|\alpha'\rangle = \hat {\bar d} | \alpha\rangle/\mathcal{E}_\alpha$, which follows from $|\alpha\rangle = \hat H|\alpha\rangle/\mathcal{E}_\alpha$, $\hat H = \hat d \hat {\bar d} + \hat {\bar d} \hat d$ and $\hat d |\alpha\rangle = 0$. Again, we have a BF pair of the same eigenvalue: $|\alpha\rangle = \hat d |\alpha'\rangle$ and $|\alpha'\rangle$. Thus, we arrived at yet another important consequence of the $\hat d$-exact evolution operator: \emph{all eigenstates with non-zero eigenvalues are non-$\hat d$-symmetric BF pairs, while all $\hat d$-symmetric eigenstates have zero eigenvalues.}

One of the ways to define the supersymmetric eigenstates is via the following property of its bra and ket: $\langle \theta|\hat d = 0$ and $\hat d|\theta\rangle=0$. It follows immediately from this definition that the expectation value on the supersymmetric eigenstates is zero not only for the stochastic evolution operator but also for any $\hat d$-exact operator. This definition also suggests that the ket (and the same with bra) of a supersymmetric eigenstate is non-trivial in the de Rahm cohomology, \emph{i.e.}, $\hat d|\theta\rangle = 0$ but $|\theta\rangle\ne \hat d|something \rangle$. The second inequality holds because if it did not, the norm of the eigenstate would vanish, $\langle\theta|\theta\rangle = \langle\theta|\hat d|\text{something}\rangle=0$.

The possible spectra of the stochastic evolution operator can be deduced from the spectral theorems for the GTO. \cite{Ruelle} These theorems ensure that for certain class of models that mimic chaotic behavior,\footnote{The models considered is Ref.\cite{Ruelle}, are "expanding" so that they have sensitivity to initial conditions thus mimicking chaotic behavior.} the GTO's eigenvalue with the largest magnitude is real. \footnote{All the other eigenvalues can be either real or come in complex conjugate pairs known as the Ruelle-Pollicott resonances (see Fig.\ref{Figure1}d) as follows immediately from the fact that $\hat H$ is a real operator.} In terms of the spectrum of the stochastic evolution operator, $\hat H = - t^{-1}log \hat{\mathpzc{M}}_t $, this means that the ground state eigenvalue is real, $\Gamma_g = min_n\Gamma_n$ (see Fig.\ref{Figure1}f-h). Let us consider from now on only models of these types. \footnote{In situations when one of the Ruelle-Pollicott resonances is the ground state, the so-called psuedo-time-reversal symmetry of pseudo-Hermitian stochastic evolution operator must also be spontaneously broken. This mechanism of the time-reversal symmetry breaking in dynamical systems can be interesting from the point of view of the search of such mechanisms (see, \emph{e.g.}, Ref.\cite{Somsikov}). Even
though we do not consider these situations here, such spectra are realizable as has been established
recently in Ref. \cite{Torsten}}

For models with compact phase spaces, each de Rahm cohomology class must provide one supersymmetric eigenstate. Otherwise, the eigensystem of the pseudo-Hermitian stochastic evolution operator would not be complete on $\Omega(X)$ because a state from any de Rahm cohomology class cannot be given as a superposition of states from other de Rahm cohomology classes and non-supersymmetric states. Another class of models that always has at least one supersymmetric state can be called "physical" models. In these models, at least one supersymmetric eigenstate from $\Omega^{D}$ exist. The ket of this eigenstate can be recognized as the steady-state (zero-eigenvalue) total probability distribution of the thermodynamic equilibrium.\footnote{In the dynamical system theory, this supersymmetric state is often called ergodic zero.} Let us further narrow our interest to the models in which at least one supersymmetric eigenstate exist and since all supersymmetric states have zero eigenvalue the spectrum Fig.\ref{Figure1}f is not realizable. 

Allow us to digress now on a brief discussion of the one of the most fundamental objects in the theory - the Witten index:
\begin{eqnarray}
W_t = {Tr} ((-1)^{\hat k}e^{-t\hat H}) = \sum_\alpha (-1)^{k_\alpha}e^{-t{\mathcal E}_\alpha} = \#(\text{bosonic }\theta's) - \#(\text{ferminic }\theta's).
\end{eqnarray}
Here, $\hat k=\chi^i\partial/\partial \chi^i$ is the operator of the number of fermions. This operator commutes with $\hat H$ and thus is a good quantum number, $\hat k|\alpha\rangle=k_\alpha|\alpha\rangle$. Only the $\hat d$-symmetric eigenstates, $\theta$'s, contribute to the Witten index because the contributions from the non-supersymmetric partners cancel each other as they have the same eigenvalues while the number of fermions differs by one. This means in particular that the Witten index is independent of time of the evolution. In fact, the Witten index is a topological object. Its physical meaning is the partition function of the stochastic noise, $\langle1\rangle_{noise}=1$, multiplied a topological constant, which for closed phase spaces equals the Euler characteristic\footnote{The Euler characteristic is one of the most fundamental topological characteristics of manifolds. For two-dimensional closed orientable topological manifolds, the Euler characteristic equals $2(1-g)$, where $g$ is the genius, \emph{e.g.}, the number of handles that one must atttach to a sphere in order to construct this manifold.} of the phase space.\cite{Mine1} The existence of the representative of the partition function of the noise as one of the fundamental objects in the theory can be viewed as a sanity check for the theory. Note that if we use the standard picture of the SDEs and viewed only the top differential forms (the total probability distributions) as the Hilbert space of the model, then this representative of the partition function of the noise would not exist at all so that the theory would not be complete. This is a very convincing proof of that the Hilbert space of SDEs is the entire exterior algebra.

Let us get back to the two remaining forms of the spectrum in Figs. 1g and 1h. The one with negative $\Gamma_g$ corresponds to the spontaneously broken topological supersymmetry, when the ground state is non-$\hat d$-symmetric because as we discussed previously all the eigenstates with non-zero eigenvalues are non-$\hat d$-symmetric. Let us recall now that the approximate meaning of the dynamical partition function,
\begin{eqnarray}
Z_t = {Tr}(e^{-t \hat H}),
\end{eqnarray}
in the large time limit, $t\to\infty$, and under certain conditions\footnote{The exact conditions when this is true are not known. What is known, however, is that the rate of growth of the stochastically averaged number of periodic solutions of SDE grows faster that $Z$. This is already enough to support the claim of this paragraph.} is the stochastically averaged number of periodic solutions, \cite{Mine1} which grows exponentially when the $\hat d$-symmetry is broken spontaneously:
\begin{eqnarray}
\left.Z_t \right|_{t\to\infty}\approx \langle \#\{\text{periodic solutions}\} \rangle_{noise} \approx e^{ t |\Gamma_g| }.\label{Growth}
\end{eqnarray}
This growth is a definitive feature of deterministic chaos and the rate of the growth is related to the various versions of  the concept of entropy in the DS. \cite{ErgodicTheoryOfChaos} Eq.(\ref{Growth}) is the stochastic generalization of this situation. Thus we come to the conclusion that the dynamical phenomenon of chaos is indeed the spontaneous breakdown of $\hat d$-symmetry. This is the universal definition of chaos that works for deterministic and stochastic models, as well as for Hamiltonian, Langevin (gradient) or more general flow vector fields. This definition has a potential to provide a unified framework for previous approaches to various realization of chaotic behavior.\cite{Zaslavsky,Kadomtsev}  

Using now the standard field-theoretic tools such as those related to the Goldstone theorem one can show that models with spontaneously broken $\hat d$-symmetry must always exhibit a long range behavior one of which is the butterfly effect.\cite{Mine2} In the continuous-space models, the associated Goldstone mode is a branch of gapless fermions (the topological supersymmetry is a fermionic symmetry) that are often called goldstinos. It is these goldstinos that due to their gaplessness bear the chaotic long-range memory/correlations in the model.

The emergence of the dynamical long-range order is yet another clear indication (the above exponential growth of periodic solutions is the other one) on that the spontaneous breakdown of topological supersymmetry must indeed be associated with the concept of stochastic chaos. 

\section{Relation to Various Physical Concepts}

The difficulties in the physical interpretation of the theory emanate from two facts. First, the ground state of a chaotic DS is not stationary/invariant in time, {\emph{i.e.}, the ground state has a non-zero eigenvalue. Second, the ground state is not a total probability distribution but a conditional probability distribution, {\emph{i.e.}}, it has a non-trivial ghost content. \footnote{In the higher-dimensional theories, this situation corresponds to the emergence of the gapless Fermi sea of ghosts (or rather of vacancies/holes for ghosts) that are called goldstinos in order to emphasize that their gaplessness is the result of the Goldstone theorem applied to the spontaneous supersymmetry breaking. In the finite-dimensional models like the ones we consider here, the Fermi sea correpsonds to the non0trivial fermionic content of the ket of the ground state, \emph{i.e.}, the ket is not a top-differential form.} This may look contradictory with the concept of thermodynamic equilibrium, which is based on the assumption that after infinitely long temporal evolution, any DS is described by a stationary total probability distribution such as the Gibbs distribution, which must be associated with (one of) its ground state(s). Another concept, which the STS picture of chaotic dynamics may seem to contradict, is ergodicity. The later suggests that the average of some observable over the infinite time is the same as the average over the "invariant measure", \emph{i.e.}, a stationary total probability distribution as a ground state.

One way to get around this contradiction is to recall that just as in quantum mechanics, it is actually the bra-ket combination that must be expected to be the total probability distribution. That the bra-ket combination is indeed the differential form of top degree, is true for all the eigenstates and not only for the ground states. Furthermore, let us now consider a vacuum expectation value (this corresponds to the physical limit of infinite temporal evolution $t\to\infty$) of an observable, $O(\hat x)$, which is a function of $x$'s only, at time moment $t>t'>0$:
\begin{eqnarray}
&\overline{O(x)}(t)=  \langle g| e^{-(t-t')\hat H} O(\hat x) e^{-t'\hat H}|g\rangle/\langle g| e^{-t\hat H}|g\rangle\nonumber \\
&= \langle g| e^{t'\mathcal{E}_g} O(\hat x) e^{- t'\mathcal{E}_g}|g\rangle = \langle g| O(\hat x) |g\rangle \nonumber \\
&= \int_{X}\bar g(x)\wedge O(x) g(x) = \int_{X} O(x) P_g(x),
\end{eqnarray}
where $P_g(x) = \bar g(x)\wedge g(x)$ is the total probability distribution associated with the ground state. As is seen, the above expectation value does not depend on time, which can be interpreted as though $P_g(x)$ is stationary. Therefore, we can always think of $P_g(x)$ as of the invariant total probability distribution required for the ergodic approach or for the thermodynamic equilibrium picture.

In the STS picture of chaos, however, it is not the total probability distribution that represents the most important, low-energy dynamics in chaotic DSs. It is the fermionic content of the ground state wavefunction that encodes the most important aspects of the low-energy chaotic dynamics. In other words, the description of a chaotic DS must go beyond that in terms of the total probability distribution. \footnote{A similar situation appears in the theory of glasses, which by the way, are believed to be chaotic or intermittent DSs. There, the conventional statistical picture is complemented by additional/fictitious degrees of freedom through the so-called replica trick. \cite{ReplicaSymmetryBreaking}} The situation is somewhat similar to quantum mechanics, in which it is the phase of the wavefunction and not its amplitude that separates quantum mechanics from a purely probabilistic description. Furthermore, it is the phase that actually determines the quantum properties of matter such as interference, diffraction etc.

The same problem can be addressed from yet another angle. Chaotic DSs can be looked upon as those out of their thermal equilibrium. They can not "thermalize" some of their variables, so that in the deterministic limit and under certain conditions they must correspond to the unstable directions with positive Lyapunov exponents. In those variables the ground state is not a "distribution" (has no ghosts/differentials). In order to make sense out of such a ground state, additional "external" knowledge is needed: something or someone must know with certainty the values of those variables at a given moment of time. The question is what or who is the bearer of this information?

The most likely answer to this question is the one that follows naturally from our previous discussion of the "bra-ket" total probability distribution. This answer is that it is the bra of the ground state that contains the information about unstable variables. The bra, in turn, represents the infinite future of the DS. \footnote{In the pathintegral language, the bra of the ground state is the functional integral over all the trajectories connecting a given (zero) time and positive temporal infinity - the infinite future of the DS.} In other words, it is the DS itself that knows these variables, but it does so not at the moment of observation but in the infinite future. Putting it differently, the DS needs yet another infinite portion of time (on top of that temporal infinity that is needed to form the ket of the ground state) in order to thermalize those variables. This picture is one of the ways to understand the chaotic long-term memory of the infinite past, \emph{e.g.}, of initial conditions.

Another possible way to answer the above question is to believe that it is the external observer that knows with certainty the values of unthermalized variables of the chaotic DS at the moment of observation. In fact, the external observer is a central figure in several physical theories. In quantum theory, \emph{e.g.}, the concept of external observer is brought to its extreme: even the mere act of observation of a quantum system leads to a wavefunction change/collapse. This brings forth yet another question about the interpretation of the theory we are dealing with. What happens when an external observer learns a value of some of the variables of a DS? The conditional probability distribution interpretation of the wavefunctions suggests that if the observation does act nontrivially, it acts most likely on the ghost sector of the wavefunction.

\section{Conclusion}

In conclusion, the supersymmetric theory of SDEs is free of approximations and for this sole reason is technically solid. What this theory is still missing, however, is a firm interpretational basis. In this Letter, we made a few important steps toward the construction of such a basis. First, we physically justified the existence of topological supersymmetry in all SDEs by connecting it to the concept of boundary that must be respected by smooth dynamics. Second, we proposed that the physical meaning of the situation when the ground state of a chaotic DS is not a total but conditional probability distribution is the failure of the DS to thermalize some of its variables and, as a result, to reach thermodynamic equilibrium. We argued that knowledge about these variables is carried by either the DSs itself but in the infinite future (the bra of the ground state) or by an external observer.

We believe that further work on the STS and, in particular, on the construction of its interpretational basis can be very insightful and probably even shed some additional light on the foundations of statistical physics and thermodynamics. After this work is done, it might become possible to construct meaningful low-energy effective theories of such dynamical systems as turbulent water or collective neurodynamical activity in brain.

\section*{Acknowledgments}

The work was supported by DARPA. One of us (KLW) would like to acknowledge the endowed support of Raytheon Chair Professorship.


\begin{thebibliography}{0}
\bibitem{Grindel} M. Ringel and V. Gritsev, Phys. Rev. A 88 (2013) 062105.
\bibitem{Kapitaniak} T. Kapitaniak, Chaos in systems with noise, World Scientific, Singapore New Jersey London Hong Kong, 1990.
\bibitem{review} A. E. Motter, and D. K. Campbell, Physics Today 66 (2013) 27.
\bibitem{Rue14} D. Ruelle, Early chaos theory, Physics Today 67 (2014) 9.
\bibitem{Univ_Chao} P. Cvitanovic, Universality in Chaos, Taylor \& Francis (1989).
\bibitem{ChaosTopology} R. Gilmore, Rev. Mod. Phys. 70 (1998) 1455.
\bibitem{brain} J. M. Beggs and D. Plenz, J. of Neuros. 23 (2003) 11167.
\bibitem{earthquakes} B. Gutenberg and C. F. Richter,  Nature 176 (1955) 795.
\bibitem{SolarFlares} M. J. Aschwanden, T. D. Tarbell, R. W. Nightingale, C. J. Schrijver, A. Title, C. C. Kankelborg, P. Martens, and H. P. Warren, Astrophys. J. 535 (2000) 1047.
\bibitem{biologicalevolution} S. J. Gould and N. Eldredge, Nature 366 (1993) 223.
\bibitem{celestialevolution} P. Garrido, S. Lovejoy, and D. Schertzer, Physica A 225 (1996) 294.
\bibitem{FinancialMarkets} T. Preis, J. J. Schneider, and H. E. Stanley, PNAS 108 (2011) 7674.
\bibitem{glasses} K. Vollmayr-Lee and E. A. Baker, Europhys. Letts. 76 (2006) 1130.
\bibitem{Nano} G.-Y. Xu, C. M. Torres, Y.-G. Zhang, F. Liu, E. B. Song, M.-S. Wang, Y. Zhou, C. Zeng, and K. Wang, Nano Letters 10 (2010) 3312.
\bibitem{Zaslavsky} G. M. Zaslavsky, Physics Reports 371 (2002) 461.
\bibitem{Kadomtsev} B B Kadomtsev, Physics-Uspekhi 38 (1995) 923.
\bibitem{ReplicaSymmetryBreaking} M. M\'ezard, G. Parisi, and M. A. Virasoro, Spin glass theory and beyond, World Scientific (Singapore, 1987).
\bibitem{Mine} I. V. Ovchinnikov, Entropy 18 (2016) 108.
\bibitem{Mine0} I. V. Ovchinnikov, Supersymmetric Theory of Stochastics: Demystification of Self-Organized Criticality, in \emph{Handbook of Applications of Chaos Theory}, Eds. C. H. Skiadas and C. Skiadas, Chapman and Hall/CRC (2016).
\bibitem{Mine2} I. V. Ovchinnikov, Transfer operators and topological field theory, arXiv:1308.4222.
\bibitem{Mine1} I. V. Ovchinnikov, Chaos 23 (2013) 013108; Chaos 22 (2012) 033134; Phys. Rev. E 83 (2011) 051129.
\bibitem{Ruelle} D. Ruelle, Notices of AMS 49 (2002) 887.
\bibitem{Somsikov} V. M. Somsikov, Open Access Library Journal 1 (2014) 1
\bibitem{WittenForms} E. Witten, J. Diff. Geom. 17 (1982) 661.
\bibitem{ParSour} G. Parisi and N. Sourlas, Phys. Rev. Letts. 43 (1979) 744.
\bibitem{MartinSR} P. C. Martin, E. D. Siggia, and H. A. Rose, Phys. Rev. A 8 (1973) 423.
\bibitem{IdentifBRST} D. Birmingham, M. Rakowski, and G. Thompson, Nucl. Phys. B 315 (1989) 577.
\bibitem{Labastida} J. M. F. Labastida, Comms. in Math. Phys.  123 (1989) 641.
\bibitem{TFTBook} D. Birmingham, M. Blau, M. Rakowski, and G. Thomson, Phys. Reps. 209 (1991) 129.
\bibitem{Kurchan} J. Tailleur , S. T\'anase-Nicola, and J. Kurchan, J. of Stat. Phys 122 (2006) 557.
\bibitem{ClassicalMechanics} E. Gozzi, Phys. Lett. B 201 (1998) 525.
\bibitem{ClassicalMechanics2} A. J. Niemi, Phys. Letts. B 355 (1995) 501.
\bibitem{Gaw} K. Gawedzki and A. Kupiainen, Nucl. Phys. B269, 45 (1986).
\bibitem{Mostafa} A. Mostafazadeh, Nucl. Phys. B 640 (2002) 419.
\bibitem{ErgodicTheoryOfChaos} J.-P. Eckmann and D. Ruelle D., Rev. Mod. Phys. 57 (1985) 617.
\bibitem{Torsten} I. V. Ovchinnikov and T. A. En{\ss}lin, {\it Kinematic Dynamo, Supersymmetry Breaking, and Chaos}, arXiv:1512.01651, accepted in Phys. Rev. D. 
\end{thebibliography}
\end{document}